# LATENT MARKOV MODEL FOR LONGITUDINAL BINARY DATA: AN APPLICATION TO THE PERFORMANCE EVALUATION OF NURSING HOMES

By Francesco Bartolucci, Monia Lupparelli and Giorgio E. Montanari

*University of Perugia, University of Bologna and University of Perugia*

Performance evaluation of nursing homes is usually accomplished by the repeated administration of questionnaires aimed at measuring the health status of the patients during their period of residence in the nursing home. We illustrate how a latent Markov model with covariates may effectively be used for the analysis of data collected in this way. This model relies on a not directly observable Markov process, whose states represent different levels of the health status. For the maximum likelihood estimation of the model we apply an EM algorithm implemented by means of certain recursions taken from the literature on hidden Markov chains. Of particular interest is the estimation of the effect of each nursing home on the probability of transition between the latent states. We show how the estimates of these effects may be used to construct a set of scores which allows us to rank these facilities in terms of their efficacy in taking care of the health conditions of their patients. The method is used within an application based on data concerning a set of nursing homes located in the Region of Umbria, Italy, which were followed for the period 2003–2005.

**1. Introduction.** Both in European countries and in the United States, elderly people with chronic conditions or functional limitations can access nursing homes whenever they are no longer able or choose not to remain in their own homes. These facilities provide a diverse array of services such as housing, support systems, nursing and medical care for a sustained period of time. These services range from minimal personal assistance to virtually total care for the patients. The challenge for the nursing homes is to provide the opportunity for elderly people to live with dignity even though they may be physically or cognitively impaired. The quality of the assistance and the efficiency of the facilities play a crucial role in restoring this sense of dignity and also in providing physical safety for the residents.

In the last decades, the increasing requirement for health assistance due to the population aging makes the quality of care in nursing homes an ever-pressing issue for policy makers. In the medical literature, there is also a great debate about the construction of indicators which measure nursing home performance and the use of these indicators to rank facilities in a certain geographical region; see Phillips et al. (2007) and the references therein. The indicators currently used to evaluate nursing home performance reflect physical conditions of elderly people and are based on data coming from surveys which are periodically carried out by public institutions; see, among others, Hirdes et al. (1998) and Mor et al. (2003). Often, a ranking of the nursing homes based on these indicators is publicly available [see Harrington et al. (2003)]. This kind of literature is strongly related to that on the evaluation of medical and health care institutions on the basis of indicators such as mortality rates. Important contributions in this literature, which are somehow related to the approach presented in this paper, are represented by the funnel plot method [see Spiegelhalter (2005)] and the hierarchical and random-effects models introduced by Normand, Glickman and Gatsonis (1997) and Ohlssen, Sharples and Spiegelhalter (2007a, 2007b).

A peculiar aspect of the surveys carried out to evaluate nursing home performance is that the same patients are usually observed at several occasions due to their long stay in the same facility. Then, we can observe how patients in each nursing home evolve in their health conditions and this is an important indicator of the performance of this facility. This aspect is not captured by the tools and the statistical models which are usually adopted for assessing nursing homes and medical care institutions.

Motivated by an application based on data coming from the Region of Umbria, Italy, in this paper we propose the use of the latent Markov (LM) model for the analysis of data on nursing homes collected by the repeated administration of questionnaires made of dichotomously-scored items. These questionnaires concern several aspects related to the health conditions of an individual. The LM model, introduced by Wiggins (1973), is a standard tool for the analysis of binary longitudinal data when the interest is in describing individual changes with respect to a certain latent status [for a review see Langeheine and van de Pol (2002)]. The latent status is represented by a latent process assumed to follow a first-order Markov chain. In the present framework, the latent status of interest is the health condition of a patient. Our approach attempts to explain: (i) how this condition changes over time depending on observable covariates and (ii) how it depends on belonging to different nursing homes. For this aim, we consider a version of the LM model where both the initial and the transition probabilities of the latent process depend on time-constant and time-varying covariates, such as gender and age [see also Vermunt, Langeheine and Böckenholt (1999)]; the model also has some connection with the mixture of experts model dealt with by Jacobs et al. (1991). Among the individual covariates, we include dummy variables for belonging to a certain nursing home. Then, the model also takes into account the multilevel structure of the data using fixed-effects, rather than random-effects, to capture the influence of each facility on the health status. This is made possible because in our application the number of nursing homes is not large. Moreover, the estimates of these effects allow us to construct a system of bidimensional scores for the performance evaluation of the nursing homes in taking care of the health conditions of their patients. Obviously, this evaluation is specially concerned with the capability of these facilities in delaying the worsening of the patient's conditions due to aging. However, these scores only provide a partial ordering for the nursing homes. Then, we also suggest a system of unidimensional scores which gives rise to a complete ordering. For the maximum likelihood estimation of the LM model illustrated in this paper we outline an EM algorithm [Dempster, Laird and Rubin (1977)] based on results well known in the hidden Markov literature [MacDonald and Zucchini (1997)] and further developed by Bartolucci (2006). We also deal in detail with model selection and the assessment of the goodness-of-fit and goodness-of-classification provided by the model.

It is worth noting that latent variable models are commonly used for the analysis of data derived from studies about living conditions; see, for instance, Mesbah, Cole and Lee (2002) and Forcina and Bartolucci (2004). However, adopting LM models in this field seems to be rather new. One of the few applications of this type is illustrated in Bartolucci, Pennoni and Lupparelli (2008), but that work is based on an LM model which is much simpler than the one dealt with in the present paper and has a different prospective from that of the performance evaluation of nursing homes. A related paper is also that of Bartolucci, Pennoni and Francis (2007) who applied a multivariate LM model to analyze a dataset based on the criminal histories of a cohort of people living in England and Wales. The aim was that of studying how the tendency to commit specific categories of crimes depends on age. Compared to the model proposed in this work, the one used in Bartolucci, Pennoni and Francis (2007) is simpler since it only allows for categorical covariates and assumes that all subjects in the sample are observed at the same occasions.

The remainder of this paper is organized as follows. Section 2 describes the dataset concerning the nursing homes located in the Region of Umbria, where the population aging is particularly evident. Section 3 describes the LM model with covariates and Section 4 describes its maximum likelihood estimation. Finally, in Section 5 we show the results of the application of the proposed approach to the dataset described in Section 2 and we conclude with a discussion in Section 6.

Table 1
*Summary statistics about the sample of 1,093 elderly people admitted in 11 nursing homes in Umbria; n.patients stands for the number of patients who were observed in the same nursing home, % males is the corresponding percentage of males and n.occasions stands for the number of occasions of administration of the questionnaire to the same patient*

| **Variable** | **Min** | **Mean** | **Max** |
|---|---|---|---|
| n.patients | 55.00 | 99.36 | 177.00 |
| % males | 20.30 | 33.58 | 40.70 |
| Age | 32.00 | 80.69 | 102.00 |
| n.occasions | 1.00 | 4.67 | 20.00 |

**2. The dataset.** In order to illustrate how the proposed approach may effectively be used to evaluate the performance of nursing homes, we consider a dataset derived from a longitudinal survey on the nursing homes operating in Umbria about the assistance level they provide to their patients. The survey is carried out since 2003 through the repeated administration of a questionnaire which is filled up by the nursing assistant of each patient and concerns several aspects of the everyday life of elderly people: cognitive conditions, ability in activity of daily living, continence self-control, disease diagnoses, skin conditions, nutritional status and the need of special treatments and medicines. In particular, we focus on the survey period 2003–2005 and consider 11 nursing homes among those located in Umbria. The resulting sample includes 1,093 residents. Summary statistics for this sample are reported in Table 1.

We also focused on a reduced set of dichotomously-scored items which are formulated so that responding 1 to any of them indicates the presence of a certain cognitive or physical limitation. This set was selected by discarding from the full questionnaire the items which: (i) do not provide any information on the physical and mental conditions which are relevant for the performance evaluation of nursing homes; (ii) have frequency of response 1 too low or too high (i.e., lower than 10% or greater than 90%). The latter follows from a standard practice in the literature on Item Response Theory [IRT; see Hambleton and Swaminathan (1985)]. Overall, we considered 9 items which are clustered in 3 groups regarding the following aspects: cognitive conditions (CC), activities of daily living (ADL) and skin conditions (SC). These items are listed in Table 2 which, for each of them, also shows the percentage of response 1 at the first occasion of administration to the same patient.

The interval of time between consecutive occasions at which the questionnaire was administered is in general equal to three months, but there are several exceptions for mainly two reasons: (i) each individual could be repeatedly charged and discharged in the same nursing home; (ii) an additional follow-up occurred whenever special treatments were required according to the patient's condition; for the same reasons, the number of these occasions is not constant across patients. This may clearly be deduced from Table 1 which shows that the number of occasions per patient ranges from 1 to 20. It is also worth mentioning that a set of personal characteristics was recorded for each patient, such as gender, date of birth, date of admission and demission and each date of administration of the questionnaire.

Table 2
*The items selected for evaluating the performance of the nursing homes. Last column shows the percentage of response* 1 *to each item at the first occasion*

| | **Item** | **%** |
|---|---|---|
| 1 | [CC1] Does the patient show problems in recalling what recently happened (5 minutes)? | 72.6 |
| 2 | [CC2] Does the patient show problems in making decisions regarding tasks of daily life? | 64.2 |
| 3 | [CC3] Does the patient have problems in being understood? | 43.9 |
| 4 | [ADL1] Does the patient need support in moving to/from lying position, turning side to side and positioning body while in bed? | 54.4 |
| 5 | [ADL2] Does the patient need support in moving to/from bed, chair, wheelchair and standing position? | 59.0 |
| 6 | [ADL3] Does the patient need support for eating? | 28.7 |
| 7 | [ADL4] Does the patient need support for using the toilet room? | 63.5 |
| 8 | [SC1] Does the patient show presence of pressure ulcers? | 15.4 |
| 9 | [SC2] Does the patient show presence of other ulcers? | 23.1 |

A preliminary assessment of the performance of the nursing homes in taking care of their patients may be based on a score assigned to each subject at each occasion of interview and defined as the percentage of item responses equal to 1. For subject $i$ observed at occasion $t$, this score is denoted by $a_{it}$ and, for the subjects in the same nursing home $h$, the evolution of the score may be summarized by the average

$$\bar{a}_h = \frac{\sum_{i:T_i>1} b_{hi} \sum_{t=1}^{T_i-1}(a_{i,t+1} - a_{it})}{\sum_{i:T_i>1} b_{hi}(T_i - 1)}.$$

In the above expression, the outer sum is extended to all subjects observed at least twice, $T_i$ is the number of response occasions for subject $i$, and $b_{hi}$ is a dummy variable equal to 1 if subject $i$ is hosted by nursing home $h$ and to 0 otherwise. A negative value of $\bar{a}_h$ means that the conditions of the patients in facility $h$ tend to improve over time, whereas a positive value means that these conditions tend to worsen. Then, these average scores allow us to rank the nursing homes according to their performance. Further information on this ranking are provided by the variability of the differential

TABLE 3
*Preliminary evaluation of the performance of each nursing home ($h$) on the basis of the average differential scores ($\bar{a}_h$) and the corresponding standard deviations ($s_h$)*

| $h$ | $\bar{a}_h$ | Rank | $s_h$ |
|---|---|---|---|
| 1 | 0.07 | (5) | 91.74 |
| 2 | 0.93 | (10) | 138.27 |
| 3 | 0.51 | (7) | 57.43 |
| 4 | −1.20 | (4) | 133.20 |
| 5 | −3.90 | (2) | 193.36 |
| 6 | −1.96 | (3) | 85.17 |
| 7 | −4.38 | (1) | 275.40 |
| 8 | 0.96 | (11) | 132.50 |
| 9 | 0.65 | (8) | 242.75 |
| 10 | 0.89 | (9) | 47.16 |
| 11 | 0.50 | (6) | 71.56 |

scores measured by the index

$$s_h = \sqrt{\frac{\sum_{i:T_i>1} b_{hi} \sum_{t=1}^{T_i-1} [(a_{i,t+1} - a_{it}) - \bar{a}_h]^2}{\sum_{i:T_i>1} b_{hi}(T_i - 1)}}.$$

The results obtained from the application of the indices $\bar{a}_h$ and $s_h$ to the available dataset are summarized in Table 3.

Two groups of nursing homes may be singled out. The first contains facilities 4, 5, 6 and 7 which have a positive effect on the health conditions of their patients. The second group contains the other facilities which, instead, admit patients whose conditions tend to worsen during time. Considering the variability of the differential scores, we can also clearly distinguish nursing homes for which this variability is low, and then their effect does not considerably vary between patients and occasions, from those for which this variability is high. Actually, a negative average differential score $\bar{a}_h$ jointly with a low standard deviation $s_h$ denotes a very good evaluation for a nursing home whose effect is positive for most of the patients. However, we can observe that the variability tends to increase as the performance improves; in fact, the highest variability is observed for facility 7, which also attains the best score.

The above analysis is only based on the observed values of the response variables and does not take into account the individual covariates and the conditions at admission and that the items may be differently related to the health status. On the other hand, the approach that will be described in the following sections takes these aspects into account. We pay particular attention to the nursing home effect in improving the health conditions of the patients or delaying the worsening of these conditions. For this aim, a nursing home may adopt special procedures, such as skin treatments for pressure ulcers, and may use special settings to support the locomotion and the ADL self-performance. We expect that the items we selected can provide an effective measure of the efficacy of these interventions.

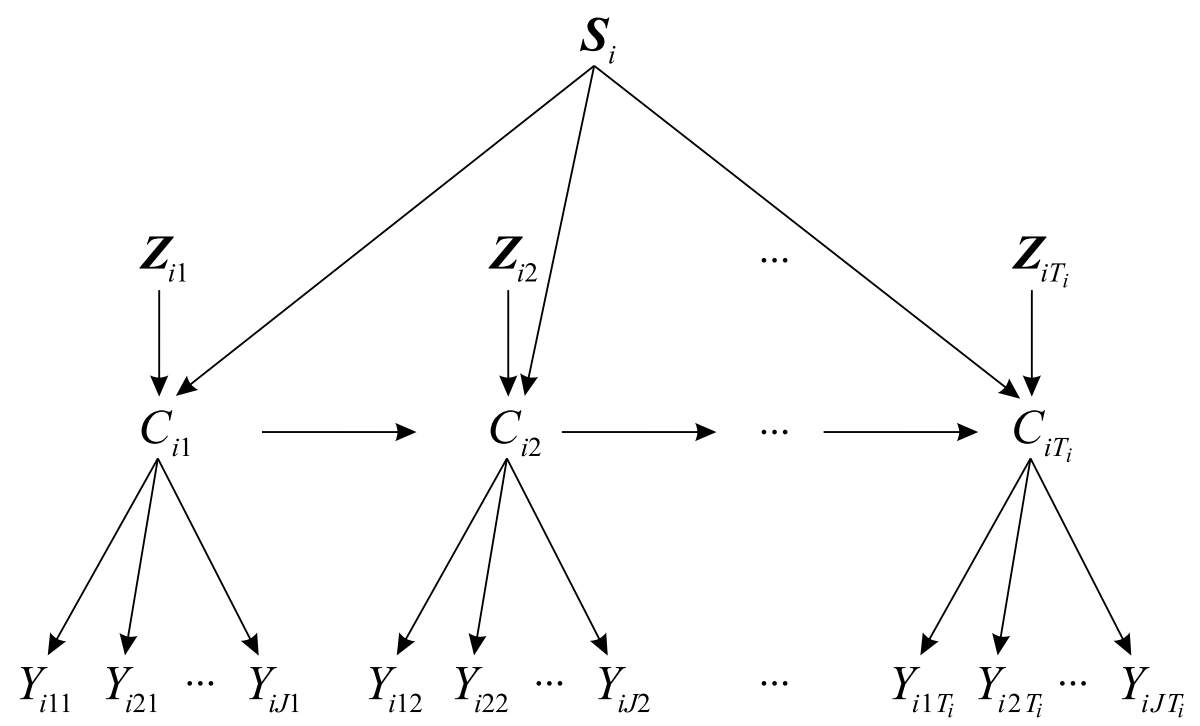

FIG. 1. *Path diagram representing the multivariate LM model with covariates for a given subject.*

**3. The latent Markov model with covariates.** Let $Y_{ijt}$ denote the binary response variable for the $j$th item administered at the $t$th occasion to the $i$th subject in the sample, with $i = 1, \ldots, n$, $j = 1, \ldots, J$ and $t = 1, \ldots, T_i$. For the same subject let $\mathbf{Y}_{it}$ be the vector with elements $Y_{ijt}$, with $j = 1, \ldots, J$, let $\mathbf{s}_i$ be the vector of time-constant covariates and let $\mathbf{z}_{it}$ be the vector of time-varying covariates, so that $\mathbf{x}_{it} = (\mathbf{s}'_i \;\; \mathbf{z}'_{it})'$ is the vector of all the covariates for this subject at occasion $t$. In our application, we take the dummy variables for coding the patient's gender and those for coding the nursing home to which the subject belongs as time-constant covariates and age and time interval between occasions as time-varying covariates.

Following an LM approach, we represent the health status of the $i$th patient at time $t$ by a discrete latent variable $C_{it}$ with $k$ support points, coded from 1 to $k$. The sequence of latent variables $C_{i1}, \ldots, C_{iT_i}$ is assumed to follow a first-order Markov chain. It is also assumed that the response variables in each vector $\mathbf{Y}_{it}$ are conditionally independent given $C_{it}$ and that the vectors $\mathbf{Y}_{i1}, \ldots, \mathbf{Y}_{iT_i}$ are conditionally independent given $C_{i1}, \ldots, C_{iT_i}$. In the latent variable literature, this assumption is usually referred to as *local independence* and, as will be clear in the following, it has important implications on the way of deriving the distribution of the response variables. Moreover, considering that our aim is the performance evaluation of nursing homes, we formulate the following assumptions on the conditional distribution of each $Y_{ijt}$ given $C_{it}$ and the initial and the transition probabilities of the latent process. The resulting model is represented by the path diagram in Figure 1.

- *Conditional distribution of $Y_{ijt}$ given $C_{it}$.* Since we want the response variables to depend only on the latent status, we assume that

$$p(Y_{ijt} = 1 | C_{it} = c, \mathbf{x}_{it}) = \lambda_j(c)$$

for each $i$, $j$, $t$ and $c = 1, \dots, k$. Moreover, we require these conditional probabilities to satisfy the constraint

$$\lambda_j(1) \le \lambda_j(2) \le \dots \le \lambda_j(k), \qquad j = 1, \dots, J, \tag{3.1}$$

so that the latent states result ordered. Since in our dataset responding 1 to an item is a sign of bad health condition, the previous constraint implies that the latent states are decreasing ordered in terms of this individual characteristic. In particular, subjects in the first state are those with the best and subjects in the last state are those with the worst health status. Note that assuming that the latent states are ordered is necessary in our context since, otherwise, it would be impossible to provide an evaluation of the nursing home effect on the probability that the health condition of a patient improves or worsens during time. In the literature on latent variable models, this is an usual assumption already adopted in similar contexts [see Forcina and Bartolucci (2004) and Bartolucci and Forcina (2005)].
- *Initial probabilities of the latent process.* Let $\pi_i(c) = p(C_{i1} = c | \mathbf{x}_{i1})$ and let $\boldsymbol{\pi}_i$ be the initial probability vector, that is, the column vector with elements $\pi_i(c)$ for $c = 1, \dots, k$. We allow these probabilities to depend on the individual covariates implementing in this way a sort of risk adjustment, that is, an adjustment for the differences between the patients at the admission in nursing homes in terms of their health conditions; for a discussion related to this point, see Normand and Shahian (2007). In particular, since the latent states are ordered, we allow these probabilities to depend on the time-constant and time-varying covariates through a parameterization which is typically adopted in regression models for ordinal variables and, in particular, in the *proportional odds model* of McCullagh (1980). We assume that

$$\log \frac{\pi_i(c+1) + \dots + \pi_i(k)}{\pi_i(1) + \dots + \pi_i(c)} = (\, \mathbf{u}_c' \quad \mathbf{x}_{i1}' \,) \boldsymbol{\beta}, \qquad i = 1, \dots, n, c = 1, \dots, k-1, \tag{3.2}$$

where $\mathbf{u}_c$ is a column vector of dimension $k-2$ with all elements equal to zero apart from the $(c-1)$th element equal to 1 when $c \ge 2$. This parametrization is based on cumulative (or global) logits which have a natural interpretation for ordinal variables. Also note that when the number of latent states increases by one, only one parameter is added to the model; this parameter corresponds to the intercept for the new state. This is because the regression coefficients for the covariates are the same for all the states and this allows the model to be parsimonious even with a large $k$.

- *Transition probabilities of the latent process.* Let $\pi_{it}(d|c) = p(C_{it} = d|C_{i,t-1} = c, \mathbf{x}_{it})$ and let $\mathbf{\Pi}_{it}$ denote the corresponding transition probability matrix with elements $\pi_{it}(d|c)$ for $c, d = 1, \dots, k$. Considering that the latent process representing the health status of an elderly patient is expected to be highly persistent, we constrain the transition matrices to be tridiagonal. With $k = 5$, for instance, we have

$$\mathbf{\Pi}_{it} = \begin{pmatrix} \pi_{it}(1|1) & \pi_{it}(2|1) & 0 & 0 & 0 \\ \pi_{it}(1|2) & \pi_{it}(2|2) & \pi_{it}(3|2) & 0 & 0 \\ 0 & \pi_{it}(2|3) & \pi_{it}(3|3) & \pi_{it}(4|3) & 0 \\ 0 & 0 & \pi_{it}(3|4) & \pi_{it}(4|4) & \pi_{it}(5|4) \\ 0 & 0 & 0 & \pi_{it}(4|5) & \pi_{it}(5|5) \end{pmatrix}.$$

This is equivalent to assuming that $\pi_{it}(d|c)$ is equal to 0 for $d \notin \{c\} \cup \mathcal{K}_c$, where $\mathcal{K}_1 = \{2\}$, $\mathcal{K}_k = \{k-1\}$ and $\mathcal{K}_c = \{c-1, c+1\}$ for $c = 2, \dots, k-1$. This has an advantage in terms of parsimony of the model, which becomes more and more evident as the number of states increases. Moreover, taking into account the individual covariates, we parameterize the transition probabilities as follows:

$$\log \frac{\pi_{it}(d|c)}{\pi_{it}(c|c)} = \mathbf{x}'_{it}\boldsymbol{\gamma}_{cd}, \qquad i = 1, \dots, n, t = 2, \dots, T_i, c = 1, \dots, k, d \in \mathcal{K}_c. \tag{3.3}$$

A version of the parametrization, which is even more parsimonious, is based on the constraints

$$\boldsymbol{\gamma}_{c,c-1} = \boldsymbol{\gamma}_1, \qquad c = 2, \dots, k, \tag{3.4}$$

and

$$\boldsymbol{\gamma}_{c,c+1} = \boldsymbol{\gamma}_2, \qquad c = 1, \dots, k-1, \tag{3.5}$$

so that the transition probabilities from state $c$ to $c-1$ and from state $c$ to $c+1$ (when these transitions are admissible) do not depend on $c$.

In the following, we indicate by M1 the model with tridiagonal transition matrices and by M2 the constrained version of M1 based on (3.4) and (3.5).

The above assumptions imply that the conditional distribution of the individual response vector $\mathbf{Y}_{it}$ given $C_{it} = c$ may be expressed as

$$m_{it}(\mathbf{y}|c) = p(\mathbf{Y}_{it} = \mathbf{y}|C_{it} = c, \mathbf{x}_{it}) = \prod_j \lambda_j(c)^{y_j}[1 - \lambda_j(c)]^{1-y_j},$$

where $\mathbf{y} = (y_1, \ldots, y_J)'$ denotes a possible realization of $\mathbf{Y}_{it}$. Moreover, the *manifest distribution* of $\mathbf{Y}_{i1}, \ldots, \mathbf{Y}_{iT_i}$ can be obtained on the basis of the factorization of the joint distribution of a first-order Markov chain as

$$\begin{aligned} q_i(\mathbf{y}_1, \ldots, \mathbf{y}_{T_i}) &= p(\mathbf{Y}_{i1} = \mathbf{y}_1, \ldots, \mathbf{Y}_{iT_i} = \mathbf{y}_{T_i} | \mathbf{x}_{i1}, \ldots, \mathbf{x}_{iT_i}) \\ &= \sum_{c_1} m_{i1}(\mathbf{y}_1|c_1)\pi_i(c_1) \sum_{c_2} m_{i2}(\mathbf{y}_2|c_2)\pi_{i2}(c_2|c_1) \cdots \\ &\quad \times \sum_{c_{T_i}} m_{iT_i}(\mathbf{y}_{T_i}|c_{T_i})\pi_{iT_i}(c_{T_i}|c_{T_i-1}). \end{aligned} \tag{3.6}$$

It has to be clear that both $m_{it}(\mathbf{y}|c)$ and $q_i(\mathbf{y}_1, \ldots, \mathbf{y}_{T_i})$ depend on the covariates. However, we consider these covariates as given and then we avoid to explicitly indicate them. A similar convention is adopted throughout the paper as, for instance, when we denote the initial probabilities by $\pi_i(c)$ and the transition probabilities by $\pi_{it}(d|c)$.

The context of application of our LM model is very different from the context of application of a hidden Markov model since the former is suitable for the analysis of data deriving from the observation of several statistical units at a limited number of occasions, whereas the latter is suitable for the analysis of one or few long series of data. However, the two models share their basic probabilistic assumptions and this implies that efficient computation of the probability in (3.6) may be performed by exploiting a forward recursion available in the hidden Markov literature [see Baum et al. (1970), Levinson, Rabiner and Sondhi (1983) and MacDonald and Zucchini (1997)]. As in Bartolucci (2006), it is convenient to express this recursion by using the matrix notation on the basis of the initial probability vectors $\boldsymbol{\pi}_i$ and transition probability matrices $\boldsymbol{\Pi}_{it}$. For this aim, consider the column vector $\mathbf{q}_{it}(\mathbf{y}_1, \ldots, \mathbf{y}_t)$ with elements $p(C_{it} = c, \mathbf{Y}_{i1} = \mathbf{y}_1, \ldots, \mathbf{Y}_{it} = \mathbf{y}_t | \mathbf{x}_{i1}, \ldots, \mathbf{x}_{it})$ for $c = 1, \ldots, k$. This vector may be computed by using the following recursion:

$$\mathbf{q}_{it}(\mathbf{y}_1, \ldots, \mathbf{y}_t) = \begin{cases} \operatorname{diag}[\mathbf{m}_{i1}(\mathbf{y}_1)]\boldsymbol{\pi}_i, & \text{if } t = 1, \\ \operatorname{diag}[\mathbf{m}_{it}(\mathbf{y}_t)]\boldsymbol{\Pi}'_{it}\mathbf{q}_{i,t-1}(\mathbf{y}_1, \ldots, \mathbf{y}_{t-1}), & \text{otherwise,} \end{cases} \tag{3.7}$$

where $\mathbf{m}_{it}(\mathbf{y}_t)$ is the column vector with elements $m_{it}(\mathbf{y}_t|c)$ for $c = 1, \ldots, k$. Once this recursion has been performed for $t = 1, \ldots, T_i$, we may obtain $q_i(\mathbf{y}_1, \ldots, \mathbf{y}_{T_i})$ as $\mathbf{q}_{iT_i}(\mathbf{y}_1, \ldots, \mathbf{y}_{T_i})'\mathbf{1}_k$, with $\mathbf{1}_k$ denoting a column vector with $k$ elements equal to 1.

An issue related to the previous one is the efficient computation of the conditional probabilities $p(C_{i,t-1} = c, C_{it} = d | \mathbf{x}_{i1}, \ldots, \mathbf{x}_{iT_i}, \mathbf{Y}_{i1} = \mathbf{y}_1, \ldots, \mathbf{Y}_{iT_i} = \mathbf{y}_{T_i})$. Let $\mathbf{R}_{it}(\mathbf{y}_1, \ldots, \mathbf{y}_{T_i})$ denote the matrix containing these probabilities for $c, d = 1, \ldots, k$. By exploiting a recursion similar to the above one, this matrix may be computed, for $t = 2, \dots, T_i$, as

$$\begin{aligned}(3.8)\quad & \mathbf{R}_{it}(\mathbf{y}_1, \dots, \mathbf{y}_{T_i}) \\ & = \frac{\operatorname{diag}[\mathbf{q}_{i,t-1}(\mathbf{y}_1, \dots, \mathbf{y}_{t-1})]\mathbf{\Pi}_{it} \operatorname{diag}[\mathbf{m}_{it}(\mathbf{y}_t)] \operatorname{diag}[\mathbf{r}_{it}(\mathbf{y}_{t+1}, \dots, \mathbf{y}_{T_i})]}{p(\mathbf{y}_1, \dots, \mathbf{y}_{T_i})},\end{aligned}$$

where the vector $\mathbf{r}_{it}(\mathbf{y}_{t+1}, \dots, \mathbf{y}_{T_i})$ is computed by the backward recursion

$$\begin{aligned}& \mathbf{r}_{it}(\mathbf{y}_{t+1}, \dots, \mathbf{y}_{T_i}) \\ & = \begin{cases} \mathbf{1}_k, & \text{if } t = T_i, \\ \mathbf{\Pi}_{i,t+1} \operatorname{diag}[\mathbf{m}_{i,t+1}(\mathbf{y}_{t+1})]\mathbf{r}_{i,t+1}(\mathbf{y}_{t+2}, \dots, \mathbf{y}_{T_i}), & \text{otherwise.} \end{cases}\end{aligned}$$

This recursion will be used to implement the estimation algorithm illustrated in the following section.

## 4. Maximum likelihood inference.

With reference to a sample of $n$ subjects, let $\mathbf{y}_{it}$ denote the observed realization of the response vector $\mathbf{Y}_{it}$, $i = 1, \dots, n$, $t = 1, \dots, T_i$. Assuming that the response vectors referred to different patients are independent given the covariates, the log-likelihood of the model illustrated above is

$$\ell(\boldsymbol{\theta}) = \sum_i \log[q_i(\mathbf{y}_{i1}, \dots, \mathbf{y}_{iT_i})],$$

where $q_i(\mathbf{y}_{i1}, \dots, \mathbf{y}_{iT_i})$ is computed by using recursion (3.7). Moreover, $\boldsymbol{\theta}$ denotes the complete parameter vector made of the subvectors $\boldsymbol{\beta}$ (for the initial probabilities of the latent process), $\boldsymbol{\gamma}$ (containing the parameters $\boldsymbol{\gamma}_{cd}$ for the transition probabilities of the latent process) and $\boldsymbol{\lambda}$ [containing the conditional probabilities $\lambda_j(c)$].

As we describe below, this log-likelihood is exploited to estimate the parameters and for model selection.

### 4.1. *Estimation.*

In order to estimate $\boldsymbol{\theta}$, we maximize $\ell(\boldsymbol{\theta})$ by an EM algorithm [Dempster, Laird and Rubin (1977)] which is based on the *complete data log-likelihood*, that is, the log-likelihood that we could compute if we knew the latent state of each subject at every occasion. This function may be expressed as

$$\ell^*(\boldsymbol{\theta}) = \sum_i \sum_c \sum_t w_{it}(c) \log[m_{it}(\mathbf{y}_{it}|c) p(C_{it} = c|\mathbf{x}_{i1}, \dots, \mathbf{x}_{it})],$$

where $w_{it}(c)$ is a dummy variable equal to 1 if subject $i$ belongs to latent state $c$ at time $t$, that is, $C_{it} = c$. The EM algorithm alternates the following two steps until convergence in $\ell(\boldsymbol{\theta})$:

- *E-step:* compute the conditional expectation of $\ell^*(\boldsymbol{\theta})$ given the observed data and the current value of $\boldsymbol{\theta}$;

- *M-step:* maximize the above expected value with respect to $\boldsymbol{\theta}$, so that this parameter vector results updated.

In order to implement these steps, it is convenient to decompose the complete data log-likelihood as $\ell^*(\boldsymbol{\theta}) = \ell_1^*(\boldsymbol{\beta}) + \ell_2^*(\boldsymbol{\gamma}) + \ell_3^*(\boldsymbol{\lambda})$ with

$$\ell_1^*(\boldsymbol{\beta}) = \sum_i \sum_c w_{i1}(c) \log[\pi_i(c)], \tag{4.1}$$

$$\ell_2^*(\boldsymbol{\gamma}) = \sum_i \sum_c \sum_d \sum_{t>1} w_{it}(c,d) \log[\pi_{it}(d|c)], \tag{4.2}$$

$$\ell_3^*(\boldsymbol{\lambda}) = \sum_i \sum_c \sum_{t>1} w_{it}(c) \log[m_{it}(\mathbf{y}_{it}|c)], \tag{4.3}$$

where $w_{it}(c,d) = w_{i,t-1}(c)w_{it}(d)$ is a dummy variable equal to 1 if subject $i$ moves from state $c$ to state $d$ at time $t$ and to 0 otherwise. The above decomposition implies that performing the E-step is equivalent to computing the conditional expected value of each dummy variable in (4.1), (4.2) and (4.3), given the observed data. Since these expected values correspond to the probabilities included in the matrices $\mathbf{R}_{it}(\mathbf{y}_{i1},\dots,\mathbf{y}_{iT_i})$, or to suitable marginalizations of these probabilities, this step may be efficiently performed by exploiting recursion (3.8).

Once the dummy variables $w_{it}(c)$ and $w_{it}(c,d)$ have been replaced by their conditional expected values, we obtain the expected value of $\ell^*(\boldsymbol{\theta})$, which is indicated by $\tilde{\ell}^*(\boldsymbol{\theta})$. The M-step updates the estimate of $\boldsymbol{\theta}$ by separately maximizing the components $\tilde{\ell}_1^*(\boldsymbol{\beta})$, $\tilde{\ell}_2^*(\boldsymbol{\gamma})$ and $\tilde{\ell}_3^*(\boldsymbol{\lambda})$ of $\tilde{\ell}^*(\boldsymbol{\theta})$, which are defined according to (4.1), (4.2) and (4.3), respectively. In particular, being based on a simple logit parameterization, $\tilde{\ell}_1^*(\boldsymbol{\beta})$ and $\tilde{\ell}_2^*(\boldsymbol{\gamma})$ are maximized by standard Newton–Raphson algorithms. On the other hand, to take into account the ordering between the probabilities $\lambda_j(c)$ defined in (3.1), maximization of $\tilde{\ell}_3^*(\boldsymbol{\lambda})$ requires a constrained version of the Newton–Raphson algorithm which may be implemented along the same lines as in Dardanoni and Forcina (1998).

We take the value at convergence of the EM algorithm as the maximum likelihood estimate of $\boldsymbol{\theta}$. This is denoted by $\hat{\boldsymbol{\theta}}$ and is made of the subvectors $\hat{\boldsymbol{\beta}}$, $\hat{\boldsymbol{\gamma}}$ and $\hat{\boldsymbol{\lambda}}$. As will be clear in Section 5, for each nursing home we compute a score on the basis of the elements of $\hat{\boldsymbol{\gamma}}$ corresponding to the dummy for being in this facility. These scores measure the effect of the nursing homes on the evolution of the health status of a patient and allow us to rank these facilities on the basis of their performance in taking care of their patients.

Finally, it is worth mentioning that the likelihood of the model described in Section 2 is typically multimodal and has a number of local maxima which increases with the number of states, but typically decreases with the sample size. The strategy that we adopt to cope with this problem is based on a preliminary exploration of the parameter space which consists of randomly selecting different points from this space and, starting from each of them, running a limited number of EM steps. Among the parameter estimates obtained in this way, the one which gives the highest likelihood is adopted to initialize the EM algorithm. Similar strategies have shown themselves successful in estimating models related to the model presented here; see, for instance, Biernacki, Celeux and Govaert (2003).

4.2. *Model selection.* A crucial point concerns the choice of the number of states of the LM model adopted in the analysis. This problem is very similar to that of the choice of the number of components of a finite mixture model, which has been deeply discussed in the statistical literature. Fundamental contributions in this sense are those of Leroux (1992) and Keribin (2000) who studied, in particular, the properties of penalized likelihood criteria. Among these criteria, that based on the Bayesian Information Criterion (BIC) seems to be preferable since, as proved by Keribin (2000), under certain conditions it leads to consistent estimation of the number of mixture components as the sample size goes to infinity. This criterion has also interesting finite sample properties; see Chapter 6 of McLachlan and Peel (2000). The use of BIC is also discussed in the hidden Markov literature and, even if its theoretical properties are not so clear, this criterion is known to perform well in choosing the number of states of a hidden Markov model; see Celeux and Durand (2008) and Boucheron and Gassiat (2007).

Taking the above considerations into account, we rely on the BIC for model selection. Using the previous notation and denoting the number of non-redundant parameters of the model of interest by $v$, we can express the index on which BIC is based as

$$\mathrm{BIC} = -2\ell(\hat{\boldsymbol{\theta}}) + v\log(n), \tag{4.4}$$

where the penalization term increases with the complexity of the model, which is measured by the number of its parameters. In particular, in order to select a model for our application, we adopt a backward strategy starting from model M1, which is the largest among the models described in Section 3. Following a standard practice, we fit this model for increasing values of $k$, the number of latent states, until we find the minimum of the BIC index. Then, with a similar criterion, we try to simplify model M1 and model M2, based on constraints (3.4) and (3.5), and then to reduce the set of covariates included into the model until it is not possible to reduce further the value of the BIC index. In doing this we always retain the number of latent states chosen under model M1. This strategy may miss the best among the available models. However, it is computationally efficient and we can expect the selected model to be reasonably close, when not identical, to the best model.

In assessing the quality of the model to be adopted, two other aspects that need to be taken into consideration are the goodness-of-fit and the goodness-of-classification. We measure the goodness-of-fit by the index

$$R^2 = 1 - \exp\{2[\hat{\ell}_0 - \ell(\hat{\boldsymbol{\theta}})]/(nJ)\},$$

where $\hat{\ell}_0$ is the maximum likelihood of the independence model, which corresponds to the proposed LM model with $k = 1$ and then has $J$ non-redundant parameters. This index may be interpreted as the average improvement of the model of interest, with respect to the independence model, in predicting each sequence of the observed responses; see also Cox and Snell (1989) and Pongsapukdee and Sukgumphaphan (2007). Similar to other indices for the goodness-of-fit of a model, $R^2$ is a relative index which ranges from 0 to 1, with higher values corresponding to a better fit. Notice that $R^2$ mainly differs from the BIC index defined in (4.4) because it does not include a term for the model complexity and then it is suitable for measuring the overall fit rather than for comparing different models.

Finally, considering that a natural criterion to classify subjects in the latent states is based on the posterior probabilities

$$p(C_{it} = c|\mathbf{x}_{i1}, \ldots, \mathbf{x}_{iT_i}, \mathbf{y}_{i1}, \ldots, \mathbf{y}_{iT_i}), \tag{4.5}$$

TABLE 4

*Results from a preliminary fitting of the LM model with different values of $k$ and different restrictions. The maximum log-likelihood of each model is denoted by $\hat{\ell}$, $v$ is the number of parameters and $k$ is the number of latent states*

| **Model** | $k$ | $v$ | $\hat{\ell}$ | **BIC** | $R^2$ | $S$ |
|---|---|---|---|---|---|---|
| M1: unrestricted LM model | 1 | 9 | −27,824 | 55,769 | – | – |
| | 2 | 59 | −18,992 | 38,397 | 0.834 | 0.989 |
| | 3 | 97 | −17,126 | 34,931 | 0.886 | 0.987 |
| | 4 | 135 | −15,880 | 32,705 | 0.912 | 0.979 |
| | 5 | 173 | −15,188 | 31,586 | 0.923 | 0.969 |
| | 6 | 211 | −14,893 | 31,262 | 0.928 | 0.967 |
| | 7 | 249 | −14,660 | 31,063 | 0.931 | 0.963 |
| | 8 | 287 | −14,568 | 31,143 | 0.932 | 0.952 |
| M2: based on restrictions (3.4) and (3.5) | 7 | 109 | −14,868 | 30,499 | 0.928 | 0.957 |
| M3: M2 + no gender effect on initial prob. | 7 | 108 | −14,870 | 30,495 | 0.928 | 0.957 |
| M4: M2 + no age effect on initial prob. | 7 | 108 | −14,888 | 30,531 | 0.928 | 0.958 |
| M5: M2 + no nursing home effect on initial prob. | 7 | 99 | −14,926 | 30,544 | 0.927 | 0.945 |
| M6: M2 + no gender effect on transition prob. | 7 | 107 | −14,870 | 30,490 | 0.928 | 0.957 |
| M7: M2 + no age effect on transition prob. | 7 | 107 | −14,870 | 30,489 | 0.928 | 0.957 |
| M8: M2 + no time effect on transition prob. | 7 | 107 | −14,885 | 30,518 | 0.928 | 0.957 |
| M9: M2 + no nursing home effect on transition prob. | 7 | 89 | −14,982 | 30,587 | 0.927 | 0.946 |
| M10: M2 + restrictions defining M3, M6, M7 | 7 | 104 | −14,875 | 30,478 | 0.928 | 0.957 |

we measure the quality of the classification by the index

$$S = \frac{\sum_{i=1}^{n}\sum_{t=1}^{T_i}(r_{it}^* - 1/k)}{(1-1/k)\sum_{i=1}^{n} T_i},$$

where, for every $i$ and $t$, $r_{it}^*$ is the maximum with respect to $c$ of the probability in (4.5). The reasoning behind this index is that when all the probabilities $r_{it}^*$ are close to 1, the classification provided by the model relies on well separated latent states. In this situation, index $S$ will attain a value close to its maximum which is equal to 1. On the other hand, when classes are not well separated, most of the probabilities $r_{it}^*$ will be close to $1/k$ and then $S$ will attain a value close to its minimum, which is equal to 0. An alternative index to measure the quality of classification could be based on the entropy, as suggested in a related context by Celeux and Soromenho (1996) and Biernacki, Celeux and Govaert (1999).

**5. Application to the dataset about nursing homes in Umbria.** In this section, we illustrate the results obtained from the application of the proposed approach to the dataset about elderly people described in Section 2. Proceeding as described in Section 4.2, we preliminary selected the number of latent states of the LM model and then we tried to simplify this model by adopting certain constraints on its parameters. We recall that we used the following covariates for modeling the initial probabilities of the latent process through (3.2): *gender* (coded by a dummy equal to 1 for a female and 0 for a male); *age* (in years); *nursing home* (coded by a suitable set of dummies). With the addition of *time between occasions* (in days), the same covariates are used to model the transition probabilities through (3.3).

For a number of latent states ($k$) between 1 and 8, Table 4 displays the maximum log-likelihood ($\hat{\ell}$) of the unrestricted LM model, indicated by M1, together with the number of parameters ($v$) and the value attained by the BIC index. The table also displays the value of the indices $R^2$ and $S$ for assessing the quality of the fit and that of the classification provided by the model. We can observe that model M1 attains the lowest value of BIC when $k = 7$. With this number of states, the model also shows a high value of $R^2$ and a very high value of $S$. On the other hand, good values of these indices are observed even with a smaller number of states, confirming the adequacy of the proposed approach for the data at hand.

We then considered several models with 7 latent states which are nested in M1. The first of these models, M2, uses only two blocks of parameters, $\boldsymbol{\gamma}_1$ and $\boldsymbol{\gamma}_2$, for the transition probabilities between latent states; see equations (3.4) and (3.5). In our application, a latent state with a smaller index corresponds to subjects in better health conditions. Then, $\hat{\boldsymbol{\gamma}}_1$ contains the parameters for the probabilities of moving to a better state and $\hat{\boldsymbol{\gamma}}_2$ contains the parameters for the probabilities of moving to a worse state. On the basis of the results in Table 4, model M2 is preferable to the unrestricted model with the same number of latent states. Proceeding in a similar way, we fitted models M3–M9 which are particular cases of M2 in which certain covariates are assumed to not affect either the initial or the transition probabilities of the latent process. By comparing these models with M2, we can conclude that gender does not have a significant effect on these probabilities. Moreover, age does not show a significant effect on the transition probabilities. The model formulated by removing these covariates, which is indicated by M10, attains the lowest BIC, equal to 30,478, among all the fitted models. Under this model, the maximum log-likelihood is equal to $-14{,}875$ with 104 parameters and we have very good values of the indices $R^2$ and $S$. Since further simplifications considerably increase the value of the BIC index, we take M10 as our final model.

The estimates of the conditional probabilities $\lambda_j(c)$ under the selected model are reported in Table 5 for each latent state $c$ and each item $j$ among those listed in Table 2. Seven ordered latent states result which represent different levels of the health status of a patient. State 1 corresponds to people in quite good conditions. States 2 and 3 correspond to patients which mainly suffer from problems related to their cognitive status. The condition of people in states 4 and 5 is aggravated by their daily activity limitations and similarly for state 6. Finally, in state 7 there are people in worst conditions because they also present problems related to the skin conditions. It may be observed that many conditional probabilities in Table 5 are equal across two or more consecutive states. This is a consequence of constraint (3.1) which is often used in the latent variable literature and, in our case, is necessary to ensure the usefulness of the results for performance evaluation. The classification of the residents in the nursing homes provided by the estimated conditional probabilities in Table 5 recalls that proposed by other authors. In particular, Kane (1998) conjectured that there exist at least five distinct groups of residents who have different needs and suffer from cognitive impairment and physical limitations at different levels. Among these groups, Kane (1998) included that of subjects who are terminally ill. In our classification, patients who are terminally ill are included among those in latent states 6 and 7.

Table 5
*Estimates of the conditional probabilities $\lambda_j(c)$ under model M10*

| **Latent state ($c$)** | **Item ($j$)** | | | | | | | | |
|---|---|---|---|---|---|---|---|---|---|
| | **1-CC1** | **2-CC2** | **3-CC3** | **4-ADL1** | **5-ADL2** | **6-ADL3** | **7-ADL4** | **8-SC1** | **9-SC2** |
| State 1 | 0.000 | 0.000 | 0.000 | 0.027 | 0.016 | 0.000 | 0.036 | 0.023 | 0.068 |
| State 2 | 0.747 | 0.000 | 0.000 | 0.027 | 0.016 | 0.011 | 0.089 | 0.026 | 0.068 |
| State 3 | 0.747 | 0.725 | 0.290 | 0.027 | 0.016 | 0.011 | 0.199 | 0.026 | 0.089 |
| State 4 | 0.747 | 0.725 | 0.290 | 0.824 | 0.950 | 0.273 | 0.968 | 0.117 | 0.224 |
| State 5 | 0.997 | 0.999 | 0.982 | 0.824 | 0.950 | 0.273 | 0.997 | 0.117 | 0.224 |
| State 6 | 0.997 | 0.999 | 0.982 | 1.000 | 1.000 | 0.927 | 0.997 | 0.117 | 0.250 |
| State 7 | 0.997 | 0.999 | 0.982 | 1.000 | 1.000 | 0.927 | 0.997 | 1.000 | 1.000 |

TABLE 6
*Estimates of the parameters (collected in $\boldsymbol{\beta}$) affecting the initial probabilities with the corresponding standard errors (s.e.), Wald test statistics (t-stat.) and p-values*

| | **Estimate** | **s.e.** | ***t*-stat.** | ***p*-value** |
|---|---|---|---|---|
| Intercept 1 | −0.393 | 0.054 | −7.306 | 0.000 |
| Intercept 2 | −1.584 | 0.093 | −16.960 | 0.000 |
| Intercept 3 | −2.865 | 0.121 | −23.765 | 0.000 |
| Intercept 4 | −3.488 | 0.131 | −26.589 | 0.000 |
| Intercept 5 | −4.731 | 0.162 | −29.188 | 0.000 |
| Age | 0.040 | 0.006 | 7.208 | 0.000 |
| Dummy 1 | −1.190 | 0.443 | −2.685 | 0.007 |
| Dummy 2 | −1.006 | 0.482 | −2.084 | 0.037 |
| Dummy 3 | −1.172 | 0.501 | −2.338 | 0.019 |
| Dummy 4 | −1.234 | 0.483 | −2.554 | 0.011 |
| Dummy 5 | −0.965 | 0.481 | −2.007 | 0.045 |
| Dummy 6 | −0.801 | 0.510 | −1.572 | 0.116 |
| Dummy 7 | −1.229 | 0.531 | −2.315 | 0.021 |
| Dummy 8 | −1.859 | 0.466 | −3.989 | 0.000 |
| Dummy 9 | −0.369 | 0.465 | −0.794 | 0.427 |
| Dummy 10 | −2.538 | 0.519 | −4.893 | 0.000 |
| Dummy 11 | −1.208 | 0.485 | −2.490 | 0.013 |

Table 6 displays the estimates of the regression coefficients for the initial probabilities of the latent Markov process, which are collected in the vector $\boldsymbol{\beta}$, together with the corresponding standard errors, Wald test statistics and $p$-values. These parameter estimates can be interpreted on the basis of assumption (3.2). In particular, there are 5 ordered intercepts, corresponding to the shift that the linear predictor has from the second to the sixth global logit, the coefficient for the covariate age and 11 coefficients for the dummies used to account for the effect of the nursing homes. Standard errors for these parameter estimates are obtained from the observed information matrix which is computed as minus the numerical derivative of the score vector. The latter may be simply obtained from the EM algorithm. Note, however, that the $p$-value associated to a dummy variable is not valid to test the hypothesis that the corresponding nursing home has an effect equal to the average effect on the health conditions of their patients. In order to test this hypothesis, a transformation of the parameter estimates similar to that we will discussed later is required (see Table 8).

TABLE 7

*Estimates of the parameters (collected in $\gamma_1$ and $\gamma_2$) affecting the transition probabilities with the corresponding standard errors (s.e.), Wald test statistics (t-stat.) and p-values*

| | **Improvement effect ($\gamma_1$)** | | | | **Worsening effect ($\gamma_2$)** | | | |
|---|---|---|---|---|---|---|---|---|
| | **Estimate** | **s.e.** | ***t*-stat.** | ***p*-value** | **Estimate** | **s.e.** | ***t*-stat.** | ***p*-value** |
| Time | −0.009 | 0.002 | −5.121 | 0.000 | −0.003 | 0.002 | −2.020 | 0.043 |
| Dummy 1 | −3.295 | 0.372 | −8.869 | 0.000 | −3.317 | 0.304 | −10.903 | 0.000 |
| Dummy 2 | −3.677 | 0.352 | −10.447 | 0.000 | −2.632 | 0.201 | −13.076 | 0.000 |
| Dummy 3 | −2.703 | 0.270 | −10.004 | 0.000 | −2.876 | 0.258 | −11.161 | 0.000 |
| Dummy 4 | −2.218 | 0.326 | −6.794 | 0.000 | −3.393 | 0.487 | −6.964 | 0.000 |
| Dummy 5 | −0.112 | 0.298 | −0.378 | 0.705 | −1.053 | 0.363 | −2.902 | 0.004 |
| Dummy 6 | −0.282 | 0.383 | −0.735 | 0.462 | −1.094 | 0.448 | −2.445 | 0.014 |
| Dummy 7 | 0.025 | 0.346 | 0.074 | 0.941 | −0.925 | 0.384 | −2.410 | 0.016 |
| Dummy 8 | −4.123 | 0.638 | −6.463 | 0.000 | −3.092 | 0.330 | −9.378 | 0.000 |
| Dummy 9 | −1.990 | 0.338 | −5.883 | 0.000 | −1.686 | 0.243 | −6.942 | 0.000 |
| Dummy 10 | −2.951 | 1.037 | −2.846 | 0.004 | −2.436 | 0.560 | −4.346 | 0.000 |
| Dummy 11 | −3.025 | 0.503 | −6.018 | 0.000 | −2.427 | 0.316 | −7.675 | 0.000 |

On the basis of the estimates in Table 6 we can conclude that older people have a greater probability, compared to younger people, to be in worse health conditions at admission in the nursing home. A certain heterogeneity between nursing homes is also observed for what concerns the type of patients they admit. For instance, nursing home 10 tends to admit patients in better health conditions, whereas nursing home 9 tends to admit patients in worse conditions. These parameter estimates correspond to the following vector of initial probabilities $\boldsymbol{\pi}_i$ averaged over all the subjects in the sample:

$$(0.133, \quad 0.048, \quad 0.217, \quad 0.282, \quad 0.111, \quad 0.134, \quad 0.073)'.$$

The latent state which, at admission, has the largest dimension is the fourth, which corresponds to subjects with some cognitive and daily activity limitations. This latent state corresponds to the 28.2% of patients and, together with the third and the fifth latent states, it amounts to more than 60% of patients. On the other hand, at the admission in the nursing home, a very low percentage of patients belongs to the latent state corresponding to the worst health conditions.

In Table 7 we show the estimates of the regression coefficients for the covariates affecting the transition probabilities of the latent Markov process which are collected in $\boldsymbol{\gamma}_1$ (improvement effect) and $\boldsymbol{\gamma}_2$ (worsening effect). In particular, each vector contains the coefficient for the time between occasions and 11 coefficients for the dummies used to account for the effect of the nursing homes. The interpretation of these parameters may be deduced from (3.3)–(3.5).

The most interesting estimates in Table 7 are those for the dummies corresponding to the nursing homes. These estimates allow us to derive a system of scores which may be used to evaluate and compare these facilities in terms of capability of taking care of the health conditions of their patients. In particular, for each nursing home $h$, let $a^*_{1h}$ denote the deviation with respect to the unweighted average of the estimate of the parameter in $\boldsymbol{\gamma}_1$ which measures the effect of this facility on the probability of improving. In a similar way we define $a^*_{2h}$ on the basis of the estimates of the parameters in $\boldsymbol{\gamma}_2$ for the probability of worsening. We then have a couple of scores $(a^*_{1h}, a^*_{2h})$, the first of which will be referred to as *improvement score* and the second will be referred to as *worsening score*. For each nursing home in the sample, the scores $(a^*_{1h}, a^*_{2h})$ are shown in Table 8 together with the corresponding standard errors and Wald test statistics and $p$-values for the hypothesis that the nursing home effect (on the probability of improving or worsening) is equal to the average effect. These scores are also represented in Figure 2 and, together with the 95% ellipsoidal confidence regions, in Figure 3.

To interpret the results we have to consider that if a nursing home has, with respect to another nursing home, a higher improvement score and a lower worsening score, the former is surely better than the latter. Therefore, each facility represented in Figure 2 is better than all the facilities located to its North West. For instance, facility 4 is surely better than facilities 1, 2, 3, 8, 10, 11. Moreover, being displayed in the fourth quadrant (see Figure 2 for the indication of the quadrant numbering), it also has a better effect than the average in taking care of the health conditions of its patients. On

TABLE 8

*Improvement and worsening scores for each nursing home (h) with the corresponding standard errors (s.e.), Wald test statistics (t-stat.) and p-values*

| | **Improvement score ($\mathbf{a}^*_1$)** | | | | **Worsening score ($\mathbf{a}^*_2$)** | | | |
|---|---|---|---|---|---|---|---|---|
| ***h*** | **Estimate** | **s.e.** | ***t*-stat.** | ***p*-value** | **Estimate** | **s.e.** | ***t*-stat.** | ***p*-value** |
| 1 | −0.789 | 0.334 | −2.363 | 0.018 | −0.909 | 0.254 | −3.585 | 0.000 |
| 2 | −1.173 | 0.340 | −3.448 | 0.001 | −0.224 | 0.193 | −1.161 | 0.246 |
| 3 | −0.198 | 0.308 | −0.642 | 0.521 | −0.468 | 0.263 | −1.779 | 0.075 |
| 4 | 0.288 | 0.320 | 0.900 | 0.368 | −0.985 | 0.442 | −2.230 | 0.026 |
| 5 | 2.393 | 0.262 | 9.147 | 0.000 | 1.355 | 0.289 | 4.687 | 0.000 |
| 6 | 2.224 | 0.352 | 6.340 | 0.000 | 1.313 | 0.366 | 3.585 | 0.000 |
| 7 | 2.531 | 0.322 | 7.867 | 0.000 | 1.483 | 0.327 | 4.535 | 0.000 |
| 8 | −1.618 | 0.529 | −3.061 | 0.002 | −0.684 | 0.255 | −2.680 | 0.007 |
| 9 | 0.515 | 0.302 | 1.707 | 0.088 | 0.722 | 0.205 | 3.522 | 0.000 |
| 10 | −0.445 | 0.975 | −0.457 | 0.648 | −0.028 | 0.495 | −0.057 | 0.955 |
| 11 | −0.520 | 0.476 | −1.092 | 0.275 | −0.019 | 0.293 | −0.066 | 0.948 |

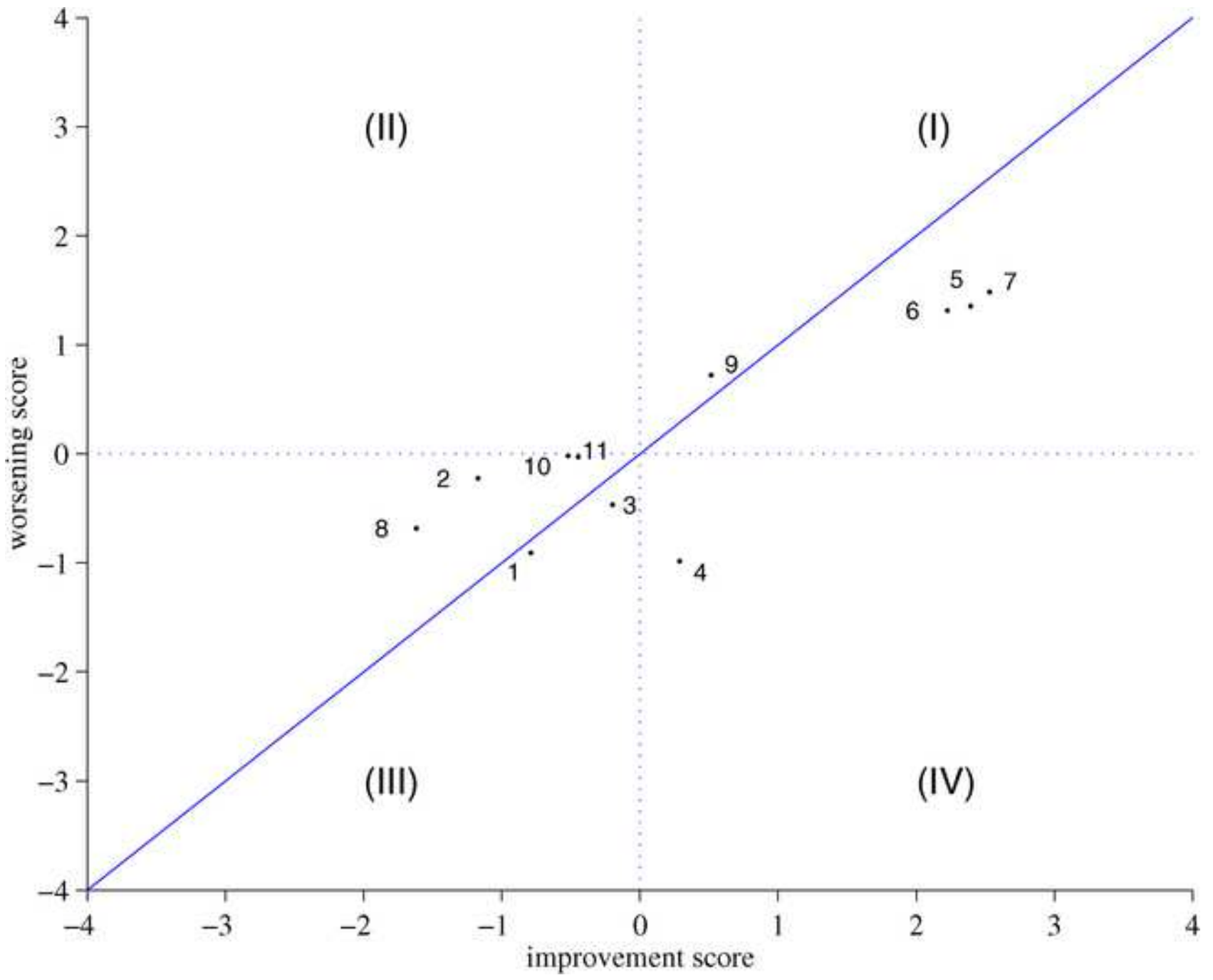

FIG. 2. *Plot of the scores* $(a^*_{1h}, a^*_{2h})$ *reported in Table 8, together with the quadrant numbering. The first score is represented on the* $x$*-axis and the second on the* $y$*-axis.*

the other hand, there are no nursing homes displayed in the second quadrant which would definitely perform worse than the average. Another aspect to take into account is that when both improvement and worsening scores are positive, a nursing home induces a lower persistence on the health condition of a patient with respect to the average. This is the case of facilities 5, 6, 7 and 9 which are displayed in the first quadrant. In contrast, patients admitted in nursing homes 1, 2 and 8, which are displayed in the third quadrant, seem to induce a higher persistence. Finally, facilities 3, 10 and 11 show a performance very close to the average effect. In fact, for these facilities, the ellipsoids in Figure 3 include the origin $(0, 0)$. These ellipsoids are also useful to assess the precision of the scores associated to each facility. For instance, we can observe that the largest ellipse is that for facility 10 for which we observe the smallest number of patients.

A drawback of the system of bidimensional scores described above is that it allows us to define just a partial ordering between the nursing homes. For instance, it is not possible to rank facilities 4 and 6 since the former has a lower improvement score, but the latter has a higher worsening score. To face this problem, we can assume that nursing homes displayed around the diagonal line in Figure 2 deserve the same evaluation because for these facilities the improvement effect is completely balanced by the worsening effect. Then, the facilities displayed below the diagonal line necessarily have a good evaluation because the positive effect on the probability of improving or the negative effect on that of worsening in the health status are predominant. In contrast, the facilities displayed over this line have a bad evaluation.

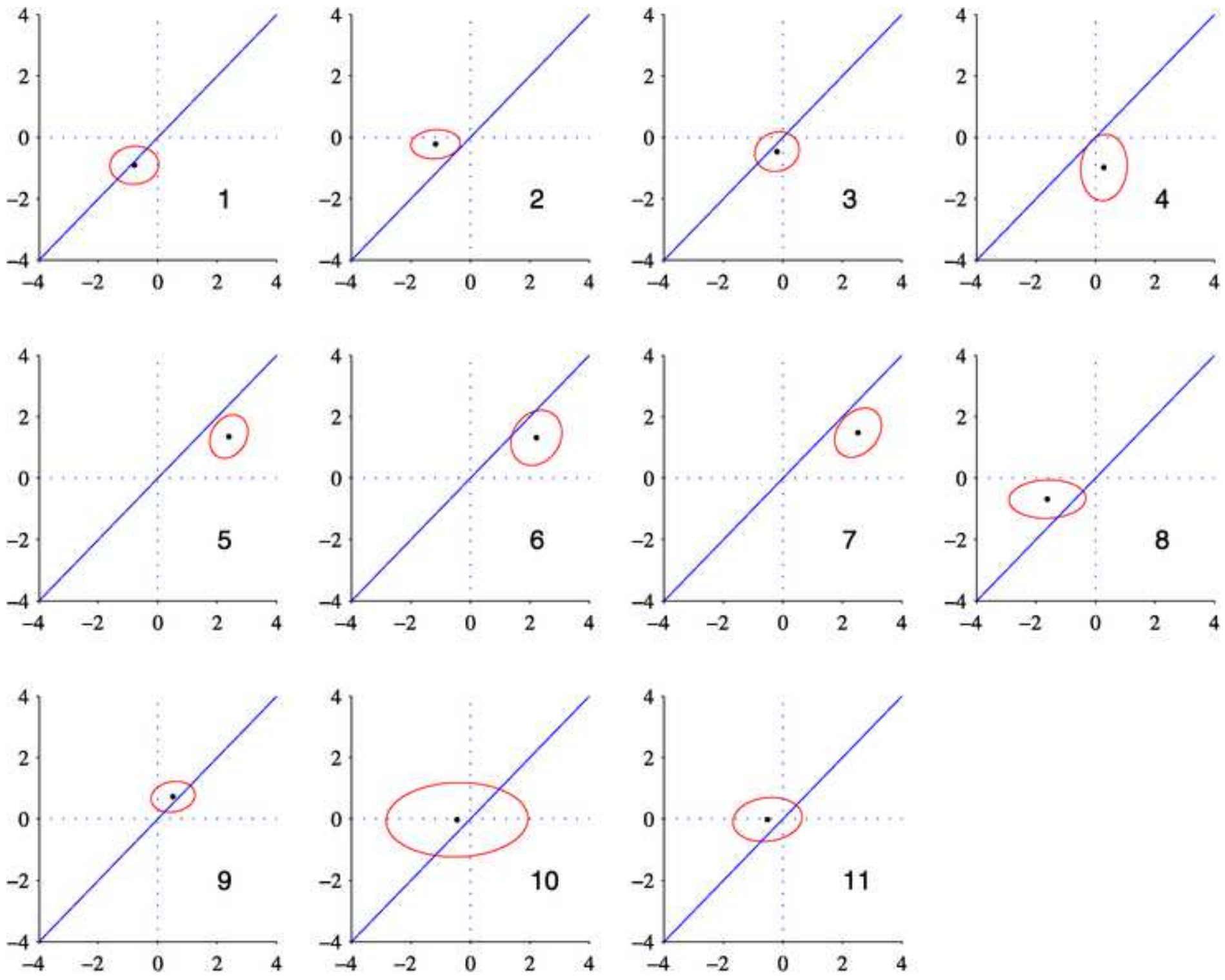

Fig. 3. *The 95% ellipsoidal confidence region for the couple of scores $(a^*_{1h}, a^*_{2h})$ of each nursing home.*

Taking into account the above arguments, we provide a system of unidimensional scores defined as $\bar{a}^*_h = a^*_{1h} - a^*_{2h}$. These scores give rise to a complete ordering of the facilities. In particular, a negative (positive) score implies a negative (positive) evaluation for the facility and corresponds to a point displayed over (below) the diagonal line in Figure 2. Moreover, the absolute value $|\bar{a}^*_h|$ is proportional to the Euclidean distance of each point $(a^*_{1h}, a^*_{2h})$ from this line. Then, the proposed system of unidimensional scores also has a geometric interpretation: a greater distance of each point displayed over (below) the diagonal line leads to a worse (better) evaluation for the nursing home. The unidimensional scores computed for the nursing homes in the sample are reported in Table 9 and represented, together with the corresponding confidence intervals, in Figure 4. In the latter, nursing homes are ordered according to the score and, then, this figure directly gives the final ranking of these facilities based on the proposed approach.

Table 9
*Unidimensional score for each nursing home (h) with the corresponding standard errors (s.e.), Wald test statistic (t-stat.) and p-value*

| | Unidimensional score ($\bar{a}^*$) | | | |
|---|---|---|---|---|
| ***h*** | **Estimate** | **s.e.** | ***t*-stat.** | ***p*-value** |
| 1 | 0.120 | 0.412 | 0.291 | 0.771 |
| 2 | −0.947 | 0.379 | −2.500 | 0.012 |
| 3 | 0.271 | 0.385 | 0.704 | 0.481 |
| 4 | 1.273 | 0.528 | 2.409 | 0.016 |
| 5 | 1.039 | 0.342 | 3.034 | 0.002 |
| 6 | 0.911 | 0.457 | 1.994 | 0.046 |
| 7 | 1.049 | 0.385 | 2.722 | 0.006 |
| 8 | −0.933 | 0.577 | −1.619 | 0.106 |
| 9 | −0.206 | 0.341 | −0.604 | 0.546 |
| 10 | −0.417 | 1.078 | −0.387 | 0.699 |
| 11 | −0.500 | 0.529 | −0.946 | 0.344 |

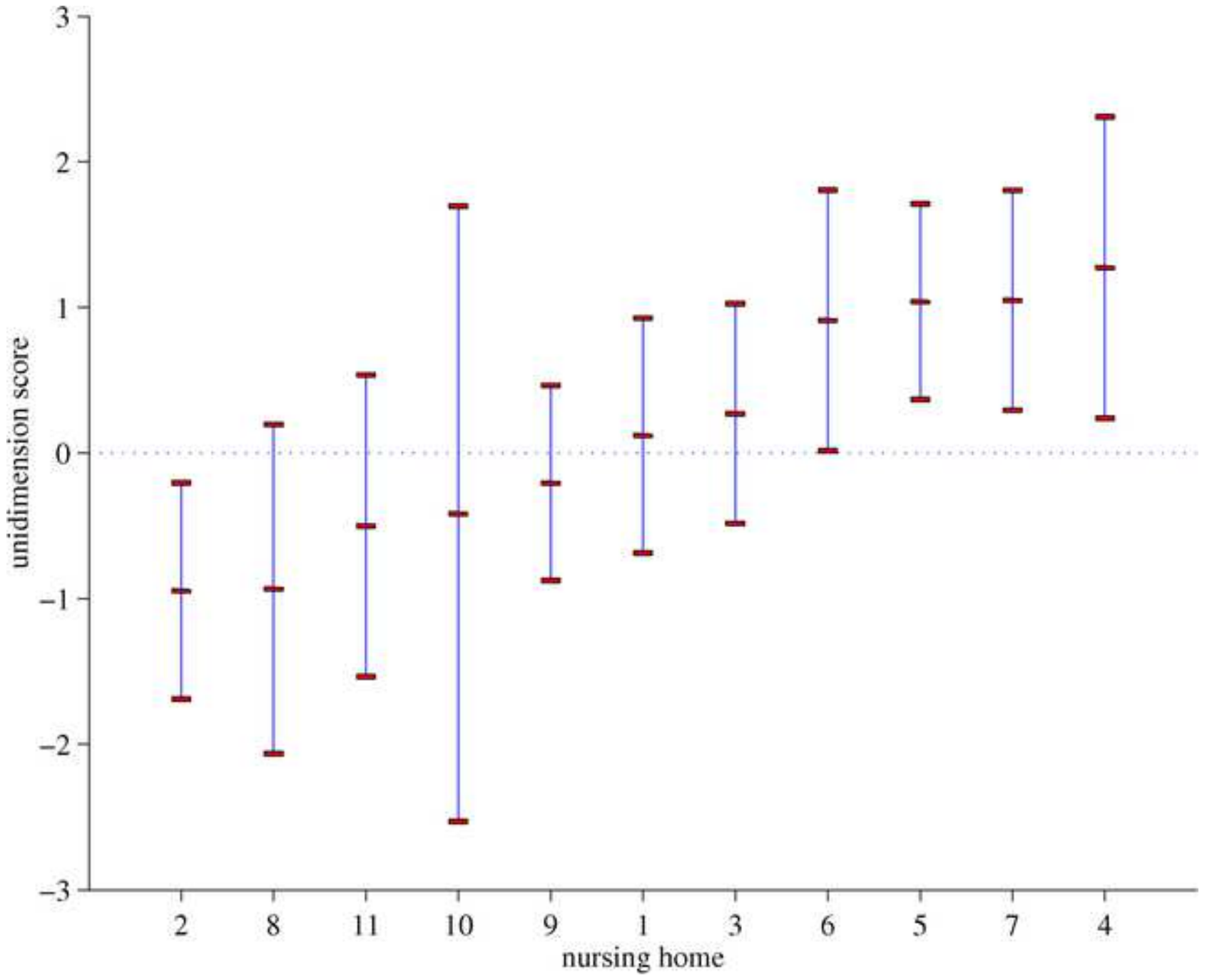

Fig. 4. *The 95% confidence interval for each unidimensional score $\bar{a}^*_h$.*

We can easily realize that nursing home 4 has the best performance in terms of taking care of the health conditions of a patient. However, its score is very close to that for nursing homes 5, 6 and 7. In fact these four facilities are displayed below the same diagonal line in Figure 2 and have a similar distance from this line. In terms of unidimensional score, these facilities are also well separated from the other ones which perform considerably worse, in particular for what concerns facilities 2 and 8. This conclusion is in partial agreement with the one reached in Section 2 on the basis of the preliminary ranking reported in Table 3. Differences are due to taking into account the admission conditions of the patients and their personal characteristics. A final comment concerns the confidence intervals shown in Figure 4. We can observe that these intervals have a similar width with the exception of that for nursing home 10 which is clearly the widest, indicating a small precision of the unidimensional score for this facility. This is in accordance with what was observed about the precision of the bidimensional scores on the basis of the ellipsoids in Figure 3.

**6. Discussion.** We proposed a latent Markov model [Wiggins (1973)] with covariates as a tool for evaluating the performance of nursing homes in taking care of the health status of their patients. This model is used for the analysis of a longitudinal dataset derived from the repeated administration of a questionnaire to a sample of patients admitted in different nursing homes in the Region of Umbria, Italy. The items contained in the questionnaire are aimed at observing several aspects related to the physical and mental conditions of elderly people in order to assess their health status. The data on which our application is based have a longitudinal structure at individual level. This is a peculiar aspect of the data used in the evaluation of nursing home performance which derives from the long stay of the patients in the same facility. Typically, this does not happen for patients of other health and medical care institutions.

By assuming the existence of a latent Markov chain for the health condition dynamics, our approach allows us to model the probability of individual changes over time. In the application, we also assumed that both the initial and the transition probabilities of the latent process depend on a set of individual covariates such as gender and age. The performance of nursing homes is evaluated by including, among the covariates, dummy variables indicating the facility hosting each patient. It is worth noting that, in contrast to other approaches for performance evaluation, our approach provides an evaluation based on how the health conditions of the patients evolve over time. Moreover, as in other approaches adopted in similar contexts [see, e.g., Normand and Shahian (2007) and Ohlssen, Sharples and Spiegelhalter (2007b)], our model has a multilevel structure which, however, is based on fixed rather than random effects. Consider also that, allowing the initial probabilities of the latent process to depend on the individual covariates, we reduce the impact of the selection bias due to possible differences between the nursing homes in terms of the characteristics of the patients they admit.

The model we adopt for performance evaluation also provides a classification of individuals in several groups corresponding to their health conditions. In our application, in particular, we found evidence of seven groups, ranging from subjects in good conditions to subjects who are cognitively impaired and have severe functional limitations. This classification is in agreement with that proposed by Kane (1998) who conjectured the existence of at least five distinct groups of residents in the nursing homes. The performance evaluation of each nursing home depends on how the facility affects the way in which its patients move between groups, and then improve or worsen in their health conditions. Our model also allows to estimate the dimension of each group and to predict the group to which every patient belongs at a given occasion and then to dynamically assess the evolution of his/her health status.

Our application mainly shows how it is possible to define a set of scores for assessing and comparing the performance of nursing homes. In particular, we discussed two different criteria. The first relies on a system of bidimiensional scores representing each nursing home effect on the probability of improving and on that of worsening in the health status. This solution also provides a graphical representation of the performance of the nursing homes (see Figure 2), but it merely defines a partial ordering between the facilities. Nursing homes belonging to the same quadrant have effects of the same direction on these probabilities and then we have a good evaluation for nursing homes displayed in the fourth quadrant since, with respect to the average, their patients have a higher probability of improving and a lower probability of worsening in their health conditions. In contrast, we have a bad evaluation for nursing homes displayed in the second quadrant. However, there is not any evident reason for preferring nursing homes in the first quadrant to those displayed in the third quadrant. Obviously, the resulting partial ordering may be difficult to use in certain situations, such as when it is necessary to decide the amount of financial support to be provided to these facilities. Thus, combining the two different effects, we proposed a system of unidimensional scores which also have a geometric interpretation. These scores may be represented in a plot (see Figure 4) which recalls similar plots adopted for the evaluation of medical care institutions on the basis of indicators such as the standardized mortality ratio; see, for instance, Spiegelhalter (2003). This solution allows us to define a complete ordering of the nursing homes in terms of their performance, which necessarily implies a certain loss of information compared to the bidimensional system. In our application based on the data collected in the Region of Umbria, through this system we identified the worst and the best nursing homes. The utility of singling out facilities with such extreme performance has been advocated by Phillips et al. (2007) as the main goal that the evaluation of nursing home performance must have.

The approach proposed in this paper can be adopted in different contexts whenever a set of items, aimed at measuring a certain latent status, is repeatedly administered to the same subjects, and these subjects are grouped according to some criteria. Consider, for instance, the problem of evaluating the health status of patients admitted in different hospitals, the productivity level of employees working in different offices, the customer satisfaction for certain products sold by different shops, and so on. However, in the presence of a large number of clusters, dummy variables for evaluating the cluster effect do not provide a parsimonious solution and a multilevel approach based on random-effects may be required. Moreover, following a standard practice in meta-analysis, a multilevel approach could be combined with an empirical Bayes approach. However, the approach based on the latent Markov model has the advantage of taking explicitly into account the dynamic nature of the health status, which is the main aspect to consider for the performance evaluation of nursing homes.

**Acknowledgments.** We acknowledge the financial support from Regione Umbria, that also provided the dataset analyzed in this paper, and MIUR (PRIN 2005—"Modelli marginali per variabili categoriche con applicazioni all'analisi causale"). See Bartolucci, Lupparelli and Montanari (2009) for the supplementary material related to the application.

## SUPPLEMENTARY MATERIAL

**Matlab functions for LM models with covariates** (DOI: 10.1214/08-AOAS230SUPP; .zip). The approach described in this paper has been implemented in a series of Matlab functions which are available from Bartolucci, Lupparelli and Montanari (2009).

F. Bartolucci
G. E. Montanari
Dipartimento di Economia
Finanza e Statistica
Università di Perugia
06123 Perugia
Italy
E-mail: bart@stat.unipg.it
giorgio@stat.unipg.it

M. Lupparelli
Dipartimento di Scienze Statistiche
Università di Bologna
40126 Bologna
Italy
E-mail: monia.lupparelli@unibo.it